\documentstyle[11pt,aaspp4,psfig]{article}
\def\lap{\lower.5ex\hbox{$\; \buildrel < \over \sim \;$}}
\def\gap{\lower.5ex\hbox{$\; \buildrel > \over \sim \;$}}
\begin{document}
\title{Mapping Low-Density Intergalactic Gas: a Third Helium Lyman-$\alpha$ 
Forest\altaffilmark{1}}
\author{ Scott F. Anderson, Craig J. Hogan, and Benjamin F. Williams}
\affil{University of Washington}
\affil{Astronomy Dept. Box 351580, Seattle, WA  98195-1580}
\affil{anderson@astro.washington.edu, hogan@astro.washington.edu, 
ben@astro.washington.edu}
\author{ Robert F. Carswell}
\affil{Institute of Astronomy}
\affil{Madingley Road, CB3 OHA Cambridge, England}
\affil{rfc@ast.cam.ac.uk}

\altaffiltext{1}{Based on observations with the NASA/ESA Hubble Space Telescope 
obtained at the Space Telescope Science Institute, which is operated by the 
Association of Universities for Research in Astronomy, Inc., under NASA
contract NAS5-26555.}

\begin{center}
Accepted for publication in The Astronomical Journal\\
{\it received 1998 Aug 11; accepted 1998 Sept 17}
\end{center}

\begin{abstract}
We present a new {\it HST}/STIS spectrum of  the $z=3.18$ quasar  PKS~1935-692 
and summarize the spectral  features shortwards of 304~\AA\ in the rest 
frame likely to be caused by foreground HeII Lyman-$\alpha$ absorption. 
In accord with previous results on two other quasars at  similar redshifts,
we demonstrate a correlation with the HI Lyman-$\alpha$ forest absorption, and
show that much of the helium absorption is caused by a  comparable quantity of 
more diffuse gas with $\Omega_g\sim 10^{-2}$, that is not detected in HI.   
The helium ionization zone around the quasar is detected as well as a void 
seen in both HI and HeII. The properties of the absorption are in broad 
agreement with those of the other quasars and with models of the protogalactic 
gas distribution and ionization at this redshift. 
\end{abstract}

\newpage
\section{Introduction}
By far the most abundant nuclei in the universe are hydrogen and helium, still 
little modified from their primordial ratio of about 12 to 1 by number. Of all 
elements  however the most abundant absorber is helium, because it is far 
less ionized than hydrogen; under the hard photoionizing conditions of 
intergalactic space at redshifts greater than 2, HeII outnumbers HI by a 
factor $\eta$ of the order of 100, and  has a greater Lyman-$\alpha$ line 
opacity  by a factor of $\eta/4\gg10$. This transition is therefore the most 
sensitive observational probe of diffuse matter. Helium opacity is  measurable 
almost everywhere---as opposed to HI absorption which over most of redshift 
space (between the HI Lyman-$\alpha$ forest lines) has a very low optical depth.
HeII Lyman-$\alpha$ absorption  occurs in the deep ultraviolet 
(303.8~\AA\  in the  rest frame, four times shorter than HI 
Lyman-$\alpha\ $), and can only be observed from space, in a small number of 
high redshift quasars which lie by chance on relatively clean lines of
sight with little foreground absorption. Three quasars are currently
known for which the HeII Lyman-$\alpha$ transition is 
accessible to {\it Hubble Space Telescope} (hereafter {\it HST})
spectroscopy, along with one more quasar at slightly lower redshift 
which has been observed by the Hopkins Ultraviolet Telescope (HUT).

Recent observational progress on HeII Lyman-$\alpha$ studies began with
Jakobsen et al.'s\markcite{Jak94} (1994) discovery of the ``HeII Gunn-Peterson effect" in 
Q0302-003 ($z=3.285$), followed closely by 
Davidsen et al.'s\markcite{David96} (1996) 
higher-resolution HUT observation of HS~1700+64 (for which HeII at $z=2.72$ 
was inaccessible to {\it HST} spectrographs), and Tytler's and Jakobsen's 
observation (Tytler \& Jakobsen\markcite{Tyt96} 1996; 
Jakobsen\markcite{Jak96} 1996) of the  
faintest of the  HeII quasars, PKS~1935-692 ($z=3.18$). These papers 
established the presence of absorption at 
low resolution and suggested the increase of the mean opacity with redshift, 
as expected in models of cosmic ionization. Later, Goddard High Resolution 
Spectrograph (GHRS) observations of Q0302-003 
(Hogan, Anderson \& Rugers\markcite{HAR97} 1997; 
hereafter HAR97) and the brighter quasar HE~2347-4342 at $z=2.885$ 
(Reimers et al.\markcite{Rei97} 1997) added enough resolution to 
begin to resolve some features of the helium Lyman-$\alpha$ forest and to 
cross-correlate with the absorption observed at the same redshift in HI, 
allowing studies of matter in the gaps between the HI forest lines. 
These data have provided the best constraints on diffuse matter at these 
redshifts, and estimates of gas density   give broad agreement with 
predictions based on numerical simulations of galaxy formation 
and extrapolations from the observed HI Lyman-$\alpha$ forest
(Croft et al.\markcite{Croft97} 1997; Zhang et al.\markcite{Zhang97} 1997; 
Zheng, Davidsen, \& Kriss\markcite{Zheng98} 1998). 
The detection of the ``proximity zone" around Q0302-003 
(HAR97\markcite{HAR97}) also allowed 
us to deploy a new set of arguments  constraining 
ionizing conditions at those redshifts, giving broad agreement with model 
expectations (Haardt \& Madau\markcite{Haardt96} 1996; Fardal, Giroux, 
\& Shull\markcite{Far98} 1998).

The introduction of the Space Telescope Imaging Spectrograph (hereafter, STIS; 
Kimble et al.\markcite{Kim98} 1998)
allows further improvements, including larger spectral range (allowing study 
at lower redshift on the same lines of sight), better background subtraction 
(allowing more precise estimates of high optical depths), and better 
sensitivity (allowing more precise optical depth estimates and in some cases 
higher spectral resolution). We have used new STIS data together with archival
{\it HST}/GHRS data  to study the  faint quasar PKS~1935-692 in enough detail 
to detect the HI correlation and the proximity zone, with comparable quality to 
the GHRS results on Q0302-003. The results here for PKS~1935-692 are remarkably 
consistent with the earlier observations of HAR97\markcite{HAR97} and 
Reimers et al.\markcite{Rei97} (1997),
supporting the idea that these few lines of sight can be used to draw general 
conclusions about the cosmic evolution of  baryon distribution and ionization.

\section{STIS Observations and Reductions for PKS~1935-692}

The new {\it HST} data discussed herein were taken on 2--3 July 1998 (UT) 
using STIS. The G140L grating and the FUV-MAMA detector provide
good throughput at modest 1.2~\AA\ spectral resolution, and cover a wide 
wavelength range of 1150--1720~\AA. The 0.2$''$ STIS slit was chosen as a 
compromise to limit geocoronal contamination, while still providing good UV 
throughput. The spectra were taken in ``TIME-TAG" mode, to eventually allow 
for even subtle comparisons of data taken during 
low vs. high background intervals, etc. However in this initial study, we 
report merely on the combined observations obtained by summing together the 
TIME-TAG observations of the July 1998 visit (i.e., our initial analysis 
here treats the STIS data as though they were simple ``ACCUM" observations). 
The spectra discussed here were processed through the standard STScI/STIS
data reduction pipeline; absolute wavelengths should be calibrated to better 
than 0.6~\AA, with the accuracy of absolute spectrophotometry better than 
$\sim$10\% (and relative wavelengths and photometry somewhat more accurate; 
e.g., see Baum et al.\markcite{Baum96} 1996).

The pipeline calibrated STIS spectrum is shown in Figure 1, overplotted with an
error spectrum which assumes $\sqrt{N}$ statistics
(errors plotted are per pixel, with 2 pixels per 
1.2 \AA\ resolution element). These STIS data present a marked improvement 
over earlier GHRS and Faint Object Spectrograph data on PKS~1935-692 for a 
number of reasons: STIS has much
lower instrumental background which is monitored simultaneously with the 
science data, excellent UV throughput, capability for efficient target 
acquisition permitting observations in a small aperture, and as a long-slit 
device much improved subtraction and monitoring of geocoronal 
contamination (e.g., compare Figures 2a and 2b). These aspects are especially 
important for PKS~1935-692, which has 
the lowest flux near HeII of any of the four ``HeII Gunn-Peterson" quasars 
discussed to date, and because geocoronal Lyman-$\alpha$ contaminates the 
spectral region of interest shortward of HeII. Figure 1 displays the grand 
average of 12,785~s of STIS exposure taken during the 5 {\it HST} orbits of the 
July 1998 observations of PKS~1935-692. The data from each individual 
{\it HST} orbit are consistent with this average spectrum, except that those 
from the second orbit may be systematically somewhat low in flux 
(perhaps suggestive of an overestimate of the background level for this one 
orbit). Pending a better understanding of the origin of this possibly 
discrepant 2nd {\it HST}-orbit pipeline reduction, we conservatively include 
it in our subsequent analysis; however, in most instances our results would 
be modified only slightly if, for example, just data from the other 4 
{\it HST} orbits (10,085~s of exposure) were considered.

\section{Results for PKS~1935-692}

Shown in Figure 2a is a portion of our STIS spectrum, highlighting
a region of interest for HeII considerations. The absorption evident below 
about 1270~\AA\ is almost surely due to HeII. The break is observed to occur 
near 1271$\pm$2~\AA\, consistent with HeII near the redshift of $z=3.185$ 
quoted for PKS~1935-692 by Jakobsen\markcite{Jak96} (1996).
\footnote{Whether or not $z=3.185$ is 
a reliable value for the ``true" cosmic or systemic redshift of the quasar 
remains to be determined as, for example, a redshift based on narrow emission 
lines has not been published for PKS~1935-692 as far as we are aware;
moreover the redshift quoted in several QSO catalogs is $z=3.17$.}
A chance superposition of a strong HI Lyman limit system causing the 
absorption break seems unlikely, as our STIS spectrum does not reveal any 
higher order HI Lyman series absorption at $z=0.4$ with HI column density 
in excess of
about 10$^{15}$--10$^{16}$cm$^{-2}$, as would likely be detectable if an 
unrelated low redshift HI Lyman limit system were the cause of the break 
near 1271~\AA. The most persuasively identified HeII absorption features are 
those matching redshifts of HI features, including the break and a prominent 
void (see below).

Blueward of the HeII break, a ``shelf" of flux is seen in PKS~1935-692 at a 
mean level of $3.4\pm 0.2\times 10^{-17}$ erg/sec/cm$^{-2}$/\AA\ 
and extends to at least about 1250~\AA\, or 20~\AA\ from the edge.
A very similar shelf was also detected in Q0302-003 by HAR97\markcite{HAR97},
and as we argued there such a shelf may plausibly be
attributed to a ``proximity effect", due to the  fact that  more of the helium 
is doubly ionized proximal to the hard radiation field of the quasar.

At 1246.5~\AA, a strong recovery void is seen in HeII that is correlated in 
redshift with a marked void in HI absorption. The reality of this recovery 
feature at 1246.5~\AA\ is hardly in doubt. First, the 
statistical significance of the feature is quite high in the 
grand average STIS spectrum. Second, the feature is strongly detected in 
STIS data from each of the 5 {\it HST} orbits, examined separately. 
Finally, this feature is also independently seen in a 41,000~s GHRS spectrum 
of PKS~1935-692 available from the {\it HST} archive (see Figure 2b), and 
therefore is surely not an instrumental artifact. (Note that the zero-level of 
the GHRS spectrum in Figure 2b is unreliable, as the GHRS data were taken using 
a $2''\times2''$ aperture and are evidently strongly contaminated by 
geocoronal Lyman-$\alpha$ and OI 1302~\AA\ airglow). 

Blueward of the proximity shelf and recovery void, the flux at $<$1242~\AA\ in 
PKS~1935-692 is further depressed, which we again attribute (see also 
Jakobsen et al.\markcite{Jak94} 1994, Davidsen et al.\markcite{David96} 1996, 
Jakobsen\markcite{Jak96} 1996, 
HAR97\markcite{HAR97}, Reimers et al.\markcite{Rei97} 1997) to absorption by 
a more diffuse intergalactic gas. For 
comparison to the results of various other studies, we estimate that the 
total HeII optical depth toward PKS~1935-692 outside the quasar proximity 
zone (and excluding the void at 1246.5\AA), including both cloud and diffuse 
contributions is $\tau_{total}\approx 3.1 (+1.4,-0.6)$ at 95\% confidence 
(but note the 99\% confidence interval includes infinite optical depth). Here 
we have compared the marginally detected mean flux of
$6.7 \pm 2.6 \times 10^{-18}$ erg/sec/cm$^2$/\AA\ averaged over the 
1225--1240~\AA\ region shortward of the shelf/void, to a representative value 
of $1.5 \times 10^{-16}$ erg/sec/cm$^2$/\AA\ longward of the break; the flux 
longward of the break is estimated from the average flux in the 
1305--1320~\AA\ region, although this value is also typical of the continuum 
level over a much broader 1300--1550~\AA\ region, as may be seen in Figure 1. 
[If we consider data from all {\it HST} 
orbits except the second, the mean flux in the 1225--1240~\AA\ region is 
$11 \times 10^{-18}$ erg/sec/cm$^2$/\AA, and 
hence $\tau_{total}\approx 2.6 (+0.8,-0.4)$.]
The value for $\tau_{total}$ estimated here for PKS~1935-692 compares well with 
that which we estimated shortward of the proximity shelf in Q0302-003 
(HAR97\markcite{HAR97}) 
of $\tau_{total}\approx 2.0 (+1.0,-0.5)$. These values are also compatible 
with the Reimers et al.\markcite{Rei97} (1997) spectrum for HE~2347-4342, 
if $\tau_{total}$ for HE~2347-4342
is also estimated over a comparably broad redshift interval.

The region blueward of geocoronal Lyman-$\alpha$ in PKS~1935-692 may potentially
prove of interest with additional data and more sophisticated reductions, but 
we do not consider it further in this initial study. The strong low-redshift HI 
Lyman-$\alpha$ feature at 1583~\AA\ evident in Figure 1 
(see Jakobsen\markcite{Jak96} 1996)
might be expected to cause strong HI Lyman limit absorption below $1190$~\AA. 
Related $z=0.3$ Lyman-$\beta$ and Lyman-$\gamma$ absorption is likely present
near 1336~\AA\ and 1266~\AA, respectively, although the former 
is probably blended with interstellar CII absorption, and the latter hard to 
interpret in detail as it falls in the proximity shelf and hence is blended 
with HeII absorption.

A more restricted portion of our new PKS~1935-692 STIS spectrum 
is displayed in Figure 3, to facilitate comparison with simple models for
the HeII absorption. Following our procedure detailed in HAR97\markcite{HAR97}, 
we overlay the observed spectrum with  a synthetic spectrum
generated from models which assume that the absorption arises
entirely from helium in the same clouds that cause the HI Lyman-$\alpha$ 
forest absorption.  In making these models, we begin with an optical 
echelle CTIO spectrum of PKS~1935-692 kindly made available by 
Outram et al.\markcite{Out98} (1998), and derive  HI line lists (from fitted  
Voigt profiles via VPFIT; Webb\markcite{Webb87} 1987). 
Then, using the HI line parameters down to a threshold 
column density of $\sim 2\times 10^{13}$~cm$^{-2}$, the HI forest cloud Doppler
parameters, redshifts, and column densities are used to predict the HeII 
absorption from the same clouds, and the predicted HeII absorption degraded to 
the STIS resolution. We also include in our models the opacity contribution 
from the strong HI system at $z=0.3$ (using an HI column density estimated from 
applying VPFIT to the STIS spectrum).

The most prominent features of 
the HeII absorption, the major HeII break itself near 1271~\AA\ and the 
marked void at 1246.5~\AA,
are accurately  predicted from the HI cloud 
models. (Note however that some of the absorption near 1266~\AA\ is 
likely related to the low-redshift strong HI Lyman-series system at $z=0.3$). 
In addition we note the hint of a correlation 
with several other observed HeII absorption 
features, in particular a tendency of the HeII flux to
recover in the voids or gaps between HI lines. 
With these new results, the correlation of some HeII features with   HI 
clouds has thus been demonstrated in all three of the HeII quasars
observed with {\it HST} resolution: Q0302-003 (HAR97\markcite{HAR97}), 
HE~2347-4342 (Reimers et al.\markcite{Rei97} 1997), and 
PKS~1935-692 (this work).

Correlating the HI and HeII absorption yields constraints
on the ionizing spectrum, cloud properties and diffuse
gas density which cannot be deduced from HI absorption alone. We derive
some specific constraints  in \S 4. 
The models displayed in Figure 3 assume two constant ratios of
HeII to HI for all the clouds, $\eta\equiv N(HeII)/N(HI) = 20$,
corresponding to a spectral slope $\alpha= 1.8$ (roughly appropriate
near the quasar), and  $\eta = 100$ (roughly appropriate to our
limit derived below far from the quasar). A change in $\eta$ indeed 
appears to be required by the data. There is, of course, some uncertainty 
even in the appropriate value to assume for the near-quasar ionizing spectral 
slope; Zheng et al.\markcite{Zheng97} (1997) find $\alpha =1.8$ for typical 
radio-quiet quasars, 
but a somewhat steeper spectral slope of $\alpha=2.2$ ($\eta \sim 35$) for 
radio-loud objects, and one might roughly estimate $\alpha \approx 2$ 
($\eta \sim 30$) from the nearby continuum data displayed in Figure 1 
for PKS~1935-692 itself. For such reasons, the models displayed in 
Figure 3 are only illustrative. These models also assume
$\xi\equiv b_{He}/b_H= 1$ corresponding
to pure ``turbulent" broadening, giving the maximal HeII/HI optical depth.
The predicted absorption depends sensitively on $\xi$, as illustrated
in Figure 4, which shows the maximal and minimal predicted
absorption for $\eta = 100$ and $\eta = 500$. 

For all the models, the   HeII absorption   
indicates the presence of  some additional gas undetected in HI,
especially in most of the voids.
Of course the gas might be detected in HI with better optical data.
Because the HI optical depth is so much smaller than that of HeII, the model 
predictions from HI magnify small observed optical depths which 
are sensitive to
noise and to systematic uncertainties such as the continuum level.
In this work we do not attempt a detailed reconstruction of these
effects but note that the HeII data require diffuse HeII absorption from 
material undetected in HI at the level of the current optical data
(corresponding to $\tau_{\rm HI}$ \lap 0.1), although the amount 
of this material required differs substantially in various portions of the 
spectrum. For example, note in Figures 3 and 4 the substantial model flux near 
1230~\AA\ that is not evident in the HeII observations. It is clear that HeII 
data provide the best information about the low-density gas at most redshifts,
and this will be true even with much better HI optical data.

\section{Implications for Diffuse Gas and Ionizing Radiation}

The properties of the absorption we find are similar in 
virtually all respects to those we found for Q0302-003, and we
interpret our results in parallel with our paper on that quasar.  
We summarize the arguments here and the new quantitative results on 
PKS~1935-692; for a more detailed discussion including the derivations, 
assumptions, and further references, see the corresponding sections (noted
below) of HAR97\markcite{HAR97}.

\subsection{Conditions Near the Quasar (HAR97 \S 3.2)}

For several arguments  concerning the absorption in the proximity 
 shelf   we wish to understand at a rough quantitative level  the influence
of the quasar on the ionization of the surrounding gas.  PKS~1935-692 is 
fainter than Q0302-003 by a factor of about 0.6, at close to  the same 
redshift, so estimates of its ionizing influence on the foreground gas scale 
by this factor. We assume that the HeII ionizing continuum flux (228~\AA) is
comparable to that longward of 304~\AA, probably a fair
assumption since the HeII Lyman-$\alpha$ emission line, where much
of the internally absorbed ionizing continuum would be reradiated,
is not strong.  However it is important to allow for the possibility of other 
sources of absorption much farther along the line of sight, such as an 
accumulation of HI Lyman continuum absorption at lower redshift than the gas 
in the proximity zone. Suppose absorption reduces the observed 304~\AA\ 
flux by a total factor $R^{-1}$; conversely, given the observed flux,  the 
flux that would be observed at the continuum edge if unabsorbed increases to
$f_\lambda\approx 1.5 \times 10^{-16}R\ {\rm erg\ cm^{-2}\ s^{-1} \AA^{-1}}$,
(or $f_\nu \approx 5 \times 10^{-30}R\ {\rm erg\ cm^{-2}\ \ s^{-1} Hz^{-1}}$).
We assume an intrinsic power law spectral energy distribution 
$f_\nu\equiv f_\lambda c\nu^{-2}\propto \nu^{-\alpha},$
where typically $\alpha\approx 1.8$.  Then, at a wavelength offset 
$\delta\lambda$ from the HeII edge we are viewing absorption
by HeII which sees an ionizing  spectral flux from the quasar,
$$
F_\nu \approx 0.7\times 10^{-23} (\delta\lambda/20\AA)^{-2}
R\ {\rm erg\ cm^{-2}\ \ s^{-1} Hz^{-1}},
$$
and the  ionizing photon flux
$$
F_\gamma\equiv \int_{\nu_i}^\infty d\nu(F_\nu/h\nu)\approx 1.1\times 10^{4}
\alpha^{-1}
(\delta\lambda/20\AA)^{-2} R {\rm cm^{-2}sec^{-1}}.
$$
These estimates are for $\Omega=1$, and differ by factors of order unity for
other cosmological models. We apply this estimate recognizing that   
uses of the quasar flux for ionization arguments in the following two
subsections will be subject to time variability of the source, so it is not 
possible to calibrate the errors in these estimates precisely.

\subsection{Quasar Lifetime and Ionizing Background from the 
Proximity Effect  (HAR97 \S 3.3)}

With this ionizing flux, the time it takes to ionize helium in a sphere
out to a distance from the quasar corresponding to spectral
offset $\delta\lambda$ is (in a flat cosmological model,
and ignoring recombinations):
$$
t_Q\ge 10^7 h^{-1} \alpha R^{-1} (\delta \lambda/ 20\AA)^3
(\Omega_gh^2/10^{-2})\ {\rm yr}.
$$
Since this is a plausible lifetime for the quasar, the shelf
can be explained as due to the second ionization of the helium
by the light of the quasar itself, even if $R$ is not much
larger than unity. 

Assume this is the case.  Then it may be that the size of the region 
influenced primarily by quasar ionization is limited by the quasar lifetime, or
that the edge is defined by the point at which the quasar light matches the 
general intergalactic background. In the latter case, we can use the proximity 
effect to estimate the background ionizing flux; in the former case, it 
gives an upper limit to the background flux. 
Taking $\delta\lambda\approx$20~\AA\ as the point where
the quasar ionizing flux is equal to the ionizing background,
the ionizing spectrum has a specific intensity
$ J_{228}\approx 0.6\times 10^{-24}R\  
{\rm erg\ cm^{-2}\ \ s^{-1} Hz^{-1}sr^{-1}}$, implying
a soft spectrum, with ratio of hydrogen to helium
intergalactic ionizing fluxes $S \equiv J_{912}/J_{228}
\approx 1.7\times 10^3 R^{-1} J_{912,-21}$. If $R$ is of
the order of unity, this large ratio implies a very soft
spectrum, consistent with the idea that breakthrough has
not yet been achieved at 228~\AA\ (e.g., Reimers et al.\markcite{Rei97} 1997);
for consistency with other estimates and models (e.g. Haardt
\& Madau\markcite{Haardt96} 1996) which give $\eta\approx 1.7 S\approx 10^2$,
we require $R\ge 10$, or an observed quasar flux smaller by a comparable
factor from its average over the last ionization-response time
($\approx 10^{6\pm1}$ years).

\subsection{Diffuse Gas Near the Quasar (HAR97 \S 3.4)}

In the region near the quasar where its flux dominates
the photoionization, we can use estimates of 
the mean optical depth $\tau_{GP}$ to estimate
the density of diffuse gas.
We use the above estimate (\S 4.1) for  the
intensity of ionizing radiation at a spectral
offset $\delta\lambda$,  tied to the observed 
quasar flux, to give the HeII/He ratio. We assume the 
standard primordial helium abundance by number 
$Y_P=0.24$ to derive a total mass density from $n(He)$.
Absorption is computed in the optically-thin limit (that is,
not counting atoms whose absorption is hidden in saturated
lines) and ignoring recombinations (a valid assumption
for low overdensity). Relaxing either of these assumptions
results in a larger density, so a lower limit to
the required overall  gas density is
$$
\Omega_g\approx 0.01 \tau_{GP}^{1/2} R^{1/2}(\alpha/1.5)^{-1/2}
(\delta\lambda/20\AA)^{-1}(h/0.7)^{1/2} 
$$

This limit may be applied to our data on the proximity shelf in the voids 
between HI clouds, where the absorption is likely dominated by underdense 
diffuse gas. We should adopt $R=1$ in foreground absorption, and an optical 
depth $\tau_{GP}$\gap 1 in the proximity zone (crudely accounting for the
discrete cloud contribution to the opacity), to derive a lower limit on
$\Omega_g$. (The limit is again
subject to uncertainty from quasar variability over $10^{6\pm1}$ year
timescales, and so cannot be precisely calibrated; to lower the limit however, 
the quasar in this case would have to have been dimmer in the past,
making the large proximity zone discussed in \S 4.2 even harder
to explain).
The allowed range for the total baryon density from nucleosynthesis and from
low redshift direct measurements  is about   
$\Omega_b\approx 0.01$ to $0.04$ (see Fukugita, Hogan, \& Peebles\markcite
{Fuk98} 1998). Thus a substantial fraction of all the baryons is
needed in diffuse, optically-thin gas to produce the observed absorption
in the near-quasar region. This also accords with 
expectations from gravitational-instability models.
Note that the ``near-quasar region" extends over  about 4,000 km/sec,
many correlation lengths, 
so the distribution  of the gas, clouds and dark matter over this
region is probably typical of the universe as whole, even
if the ionization state is not. 

The low-$\tau$ void   at 1246.5~\AA\ in PKS~1935-692 can be explained as a 
modest underdensity ($\sim 10^{-1}$) of gas,
extending over a fairly large interval 3~\AA\ $\approx$ 1000 km/sec.
This type of structure is expected on occasion.

\subsection{Diffuse Gas in the HI Lyman-$\alpha$ Voids (HAR97 \S 3.6, \S 3.9)}
We now turn to results which are pure foreground effects,
not dependent on the properties of the quasar.

First, a   conservative lower limit on the  
gas density can be derived from simply assuming that all of the 
helium is in the form of HeII. This yields 
$$
\Omega_g h^2 \ge 1.7\times 10^{-5}h
\tau_{GP}[n({\rm HeII)}/n({\rm He})][(1+z)/4.185]^{-3/2}.
$$
At any given redshift, this is the density implied 
to achieve an optical depth $\tau_{GP}$. If one
considers the integrated contribution of
 all matter then this applies regardless of saturated lines or 
thermal broadening, since it simply counts the mean
number of atoms needed to achieve a given optical depth
and this increases with any saturation. 
In the gaps far between identified lines, we know
there is still absorption (beyond the thermal wings of the lines)
and this gives the minimal amount of matter required to 
produce absorption at those redshifts.

We can reverse this argument if, as observed, some flux 
is detected far from the quasar. It  is implausible to evacuate 
gas   to   a very low density (e.g. $\Omega \le 10^{-4}$) by 
gravitational instability. If detectable  flux gets through then
over some redshift intervals     
the helium is mostly doubly ionized HeIII.
There may be unresolved gaps between HeIII regions of high opacity 
 but they can  fill at most a fraction $1- <\tau>^{-1}$ of the
volume where $<\tau>$ is the smoothed optical depth.

Note that Reimers et al.\markcite{Rei97} (1997) found evidence in HE~2347-4342
for   large (6--10 \AA) regions of the spectrum consistent with zero flux
(see Figure 5c). If zero-flux regions are large and common then the epoch of 
HeII ionization is being observed.  The fact that no definitive evidence 
appears for such regions in PKS~1935-692 (nor in GHRS spectra for Q0302-003) 
is consistent with HeII ionization being mostly finished. The differing 
results for HE~2347-4342 may be due to real variations in conditions over the 
limited spectral (redshift) regions analyzed thus far, or chance differences 
between the lines of sight, or to artifacts caused by the  imperfect control 
of backgrounds in GHRS (e.g., HAR97\markcite{HAR97} \S 2, 
Heap\markcite{Heap97} 1997). For 
example, if we re-reduce the Reimers et al. HE~2347-4342 GHRS spectrum using 
only a subset of the data taken during intervals of low noise background
(defined arbitrarily as those 13,700~s of data during
which mean GHRS background count rates were $<0.008$ counts/sec/diode), we 
find more consistency with  the spectra of the other two quasars, 
including a proximity shelf from the quasar itself (see Figure 5d).
On the other hand, such a re-reduction may also yield a biased estimate of the 
zero-level, because for GHRS (unlike STIS) background noise events 
(which occur in ``bursts") are not measured strictly simultaneously 
with the on-object counts.
The true situation should be resolved soon with additional STIS data on all 
three quasars.

\subsection{Ionizing Spectrum Far from the Quasar (HAR97 \S 3.7)}

As in Q0302-003 and HE~2347-4342, we find gaps in the HI forest lines where 
there is no recovery in the helium spectrum (e.g., see Figures 3--4
near $\sim$1230\AA). The ratios of HI and HeII optical depths in such (HI)
gaps can be used to derive the spectral hardness, without knowing the gas 
density but again assuming primordial abundances.
In a conspicuous Lyman-$\alpha$ void   (e.g., near $\lambda_{HI}
\approx 4\times 1230$~\AA), the average optical depth  of diffuse HI absorption
allowed by our current data is about 0.05. 
The optical depth of HeII is at least of order 
unity, requiring $\eta$ \gap $4\times 1\div 0.05 = 80$, in rough
agreement with, e.g., Haardt \& Madau's\markcite{Haardt96} (1996) prediction. 
The ratio can be used to
constrain the spectrum in the usual way. This limit and the following can be 
made more reliable with a better signal-to-noise spectrum of the HI forest.

\subsection{Upper Limit on $z$-Filling Gas (HAR97 \S 3.8)}

Using this limit on the spectrum, we can deduce a limit on the
IGM density, tied not to the ionizing flux from the quasar
(as we did above) but to the cosmic ionizing flux at the HI Lyman edge, 
$J_{912,-21}$
(in units of $10^{-21}{\rm erg\ cm^{-2}\ \ s^{-1} sr^{-1} Hz^{-1}}$), 
which has other observational constraints such as the HI clouds proximity 
effect. Suppose that there is a smooth density of gas at all redshifts, even
in the voids; the fact that flux gets through on average means that 
the density cannot be too high. Calculating the Gunn-Peterson optical depth 
for smooth photoionized gas yields 
$$
\Omega_g= 0.018 (\tau_{GP}/3)^{0.5}  
  ( h/0.7)^{-1.5} (\eta/100)^{-0.5}
 (J_{912,-21}/0.5)^{0.5}.
$$  
As in Q0302-003 we thus get an upper limit $\Omega_g$\lap 0.02 by using
representative values estimated above of $\eta$\gap$10^2$ and   $\tau_{GP}$\lap 3, 
and using  typical estimates $J_{912,-21}\approx 0.5$ 
(Haardt \& Madau\markcite{Haardt96} 1996).

The assumptions leading to this limit are conservative
in most respects. Modest density perturbations lead to 
more recombinations, therefore more absorption for a given
mean density. It is true that the absorption per atom is
reduced  in   highly saturated lines,
but we know that the saturated material is dominated by
the clouds already detected in the 
HI lines. Therefore, material between the identified HI forest
clouds cannot be repository for
a large number of baryons far in excess of the nucleosynthesis
value of $\Omega\approx 0.02$. Since helium is an inert gas,
this argument  applies even to models where the hydrogen is 
locked up in molecular or solid form. 

\section{Conclusions}

There is broad consistency with a picture in which some helium absorption 
occurs in the same clouds giving rise to the HI Lyman-$\alpha$ forest, but 
with additional helium absorption by material with $\Omega_g \sim 10^{-2}$ 
that is essentially similar to the HI forest clouds, albeit at a lower density 
and  filling a substantial fraction of the space between them. The total amount
of material in this latter form is comparable to, but possibly somewhat smaller 
than, that in the clouds; it is certainly not much greater. These constraints 
on the overall distribution of baryonic material cannot be derived on the basis 
of the hydrogen data alone, although they agree with the   expectations for the 
baryon density derived from nucleosynthesis, and with the
gravitational-instability formation model of the clouds. This overall picture 
is consistent for all three currently known quasars where HeII Lyman-$\alpha$
is observable with {\it HST}, and therefore may be representative of the
Universe generally at $z\approx3$.

\section{Acknowledgments}
We are grateful for the superb support to this program provided by S. Baum, 
K.~Peterson, and others at STScI. We thank J. Baldwin, G. Williger, and 
P. Outram for kindly providing access to the CTIO optical data on 
PKS~1935-692. We also thank A.~Davidsen for useful comments on this 
paper, and J.~Wadsley for interesting discussions of some theoretical aspects.
This work was
supported at the University of Washington by NASA/{\it HST} grant 
GO-07272.01.96A, and is  based  on
observations with the NASA/ESA Hubble Space Telescope, obtained  at the Space 
Telescope Science Institute, which is operated by AURA, Inc. under NASA 
contract.

\newpage

\newpage
\begin{figure}
\psfig{figure=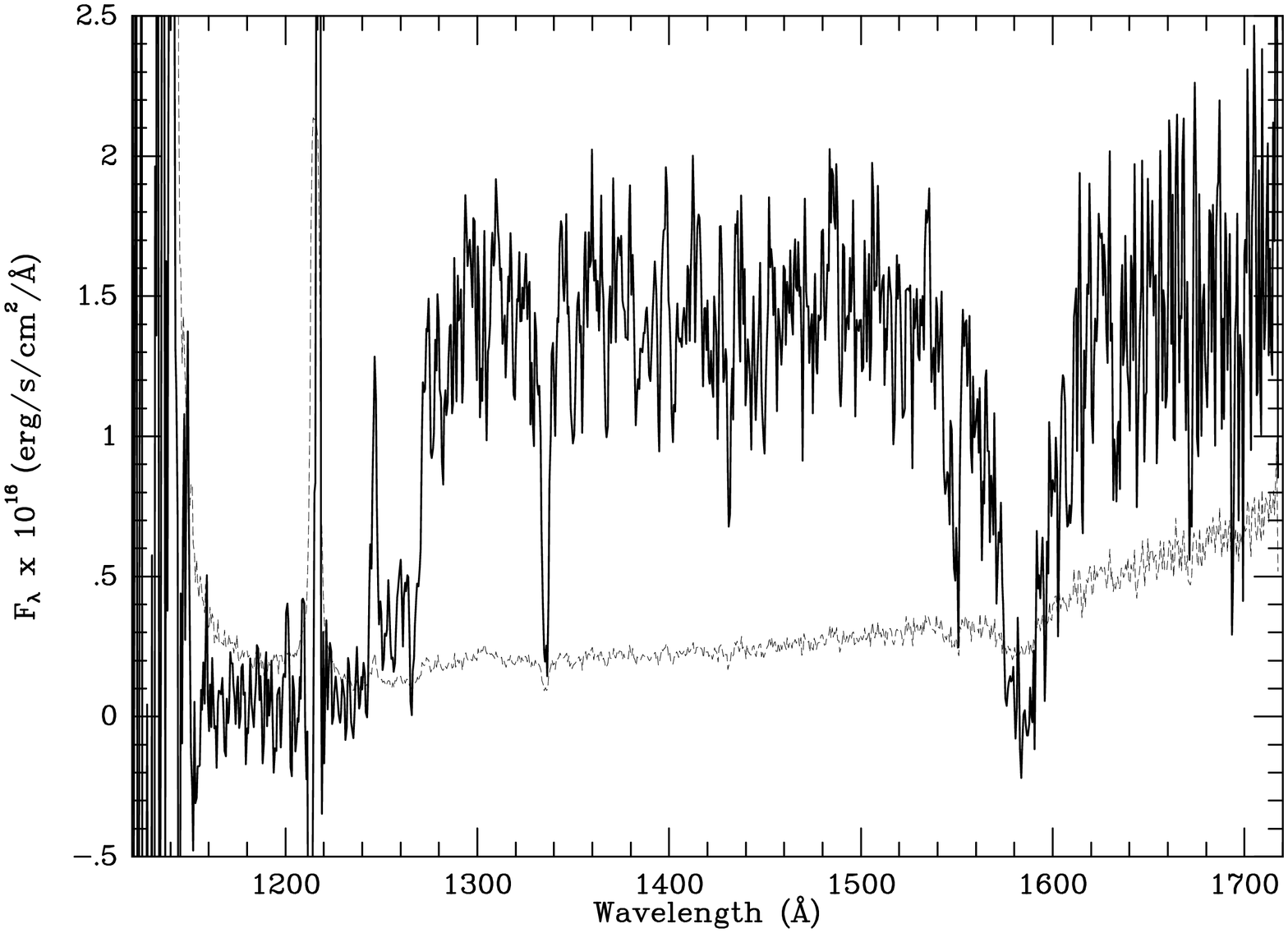,height=12truecm,width=15truecm,angle=270}
\caption[]{Average {\it HST}/STIS G140L spectrum of PKS~1935-692, and 
an error spectrum, derived from 5 {\it HST} orbits (12,785~s) of exposure 
in July 1998. (Errors plotted are per pixel, with 2 pixels per 1.2~\AA\
resolution element.)}
\end{figure}

\begin{figure}
\hskip 1.0in
\psfig{figure=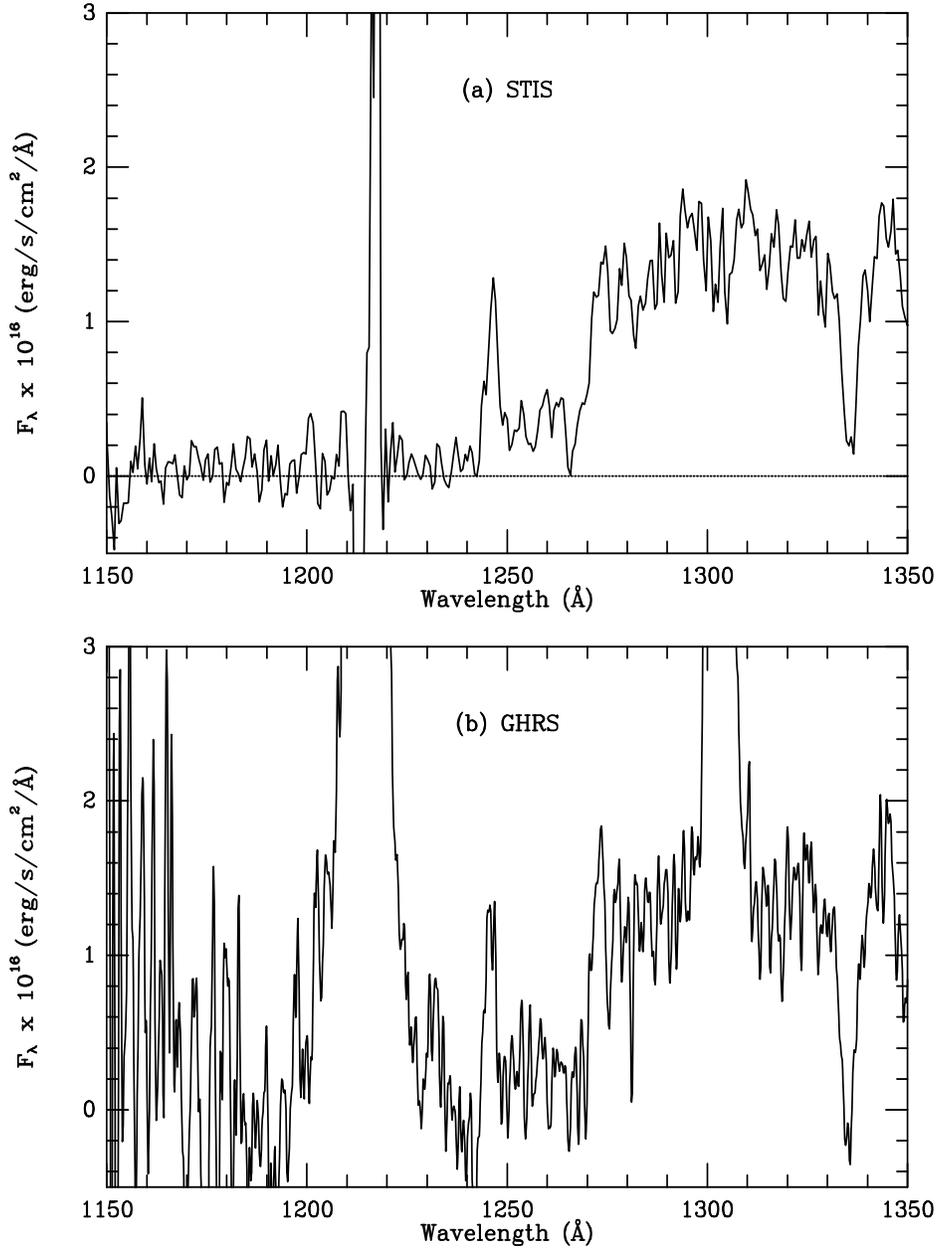,height=18truecm,width=12truecm,angle=0}
\caption[]{A portion of the {\it HST} STIS spectrum of 
PKS~1935-692, highlighting a wavelength range potentially of 
interest for HeII absorption. (a) (upper panel) Note the strong break
near 1271~\AA\ at the wavelength expected for HeII at $z=3.18$, the
proximity ``shelf" centered around 1260~\AA, the void in HeII at 1246.5~\AA, and
likely non-zero flux even in the 1225--1240~\AA\ region shortward of the
HeII break and void. The feature at 1216~\AA\ is geocoronal Lyman-$\alpha$.
(b)~(lower panel) A 41,000~s archival GHRS spectrum of PKS~1935-692. Although
the zero level is suspect due to strong contamination by
geocoronal/airglow emission (HI Lyman-$\alpha$ and OI 1302 \AA), a 
comparison of data in upper and lower panels confirms the reality of the HeII 
void at 1246.5 \AA.}
\end{figure}

\begin{figure}
\hskip 1.0in
\psfig{figure=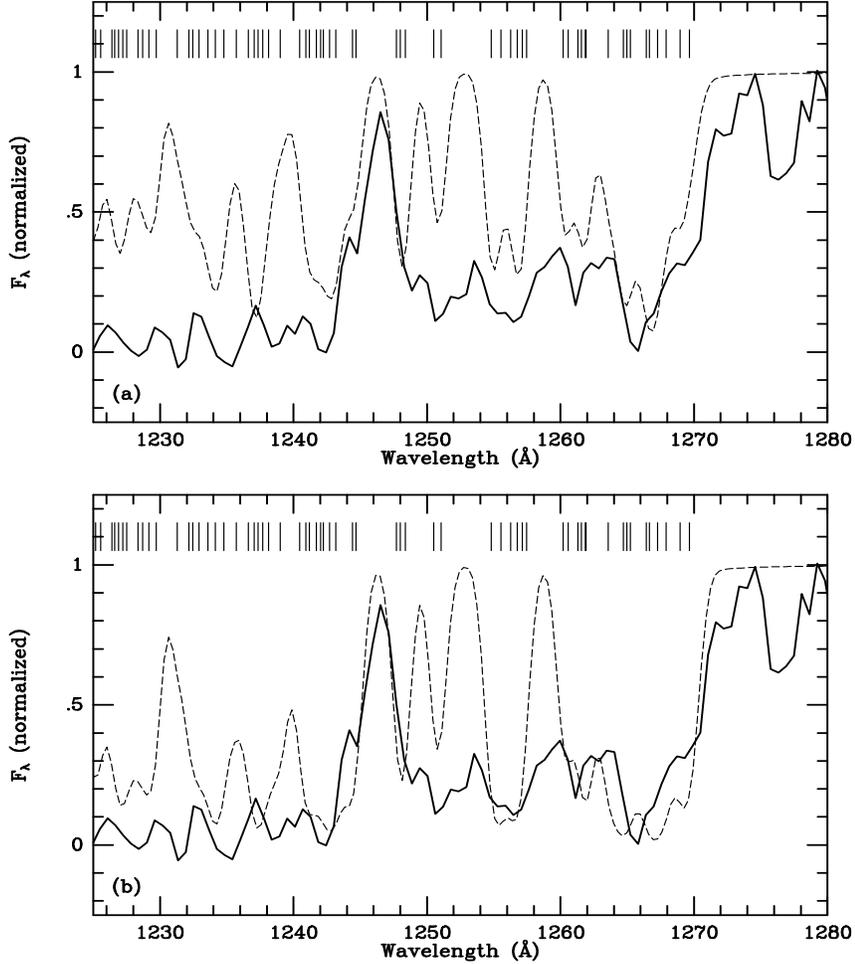,height=14truecm,width=11truecm,angle=0}
\caption[]{A more restricted portion of the {\it HST} STIS spectrum of 
PKS~1935-692, focusing on the wavelength range of most interest for HeII 
absorption. Overlaid on the observed STIS spectrum (heavy solid curve) is a 
model spectrum (dashed curve) predicted on the basis of the model  
distribution of 
HI derived from an optical spectrum of the HI Lyman-$\alpha$ forest.
The spectrum has been normalized to the continuum longward of the HeII
break, by dividing by $1.5\times 10^{-16}$erg/s/cm$^2$/\AA.
Tick marks indicate the fitted  HI velocity components from the optical
spectrum. HI Doppler parameters and column densities from the fit were
used to predict the  HeII absorption spectrum at the STIS resolution.
(a) (upper panel) HeII absorption predicted by a model with
$\eta =20$ and assuming pure turbulent broadening, $b_{HeII}=b_{HI}$.
(b) (lower panel) Plotted for comparison is the model for 
$\eta=100$ (assuming pure turbulent broadening). Note
that the models (based on HI clouds) and the STIS spectra
appear to show corresponding absorption features near the quasar,
especially including the HeII edge itself, and a void near 1246.5~\AA.
While helium in the HI forest clouds accounts for much of the opacity
in the proximity zone, there is significant HeII opacity ($\tau_{GP}$\gap 1), 
even at the redshift of the conspicuous HI Lyman-$\alpha$  forest void near 
1230~\AA. Nonzero flux is likely detected even far from the quasar, in
the 1225--1240~\AA\ regime.}
\end{figure}

\begin{figure}
\hskip 1.0in
\psfig{figure=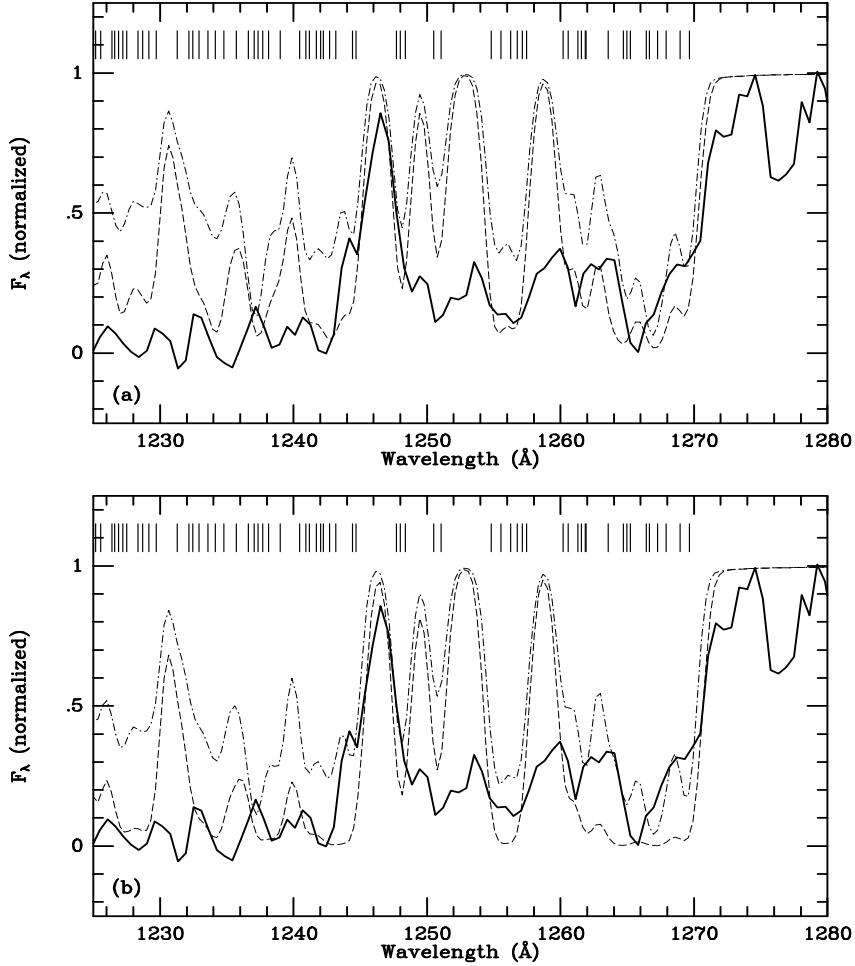,height=14truecm,width=11truecm,angle=0}
\caption[]{A restricted portion of the {\it HST} STIS spectrum of 
PKS~1935-692 (heavy solid curves), focusing on the wavelength range of most 
interest for HeII absorption, and overlaid with  absorption models based on 
the model  distribution of HI derived from an optical spectrum of the HI 
Lyman-$\alpha$ 
forest, illustrating the effect of different physical mechanisms of line 
broadening. Because the helium absorption in the line models is highly 
saturated if $\xi=0.5$ (thermal broadening), even very large  values of 
$\eta$ do not yield significantly more absorption. 
The spectrum has again been normalized to the continuum longward of the HeII
break, by dividing by $1.5\times 10^{-16}$erg/s/cm$^2$/\AA. Tick marks 
indicate the fitted  HI velocity components from the optical spectrum. 
(a) (upper panel) Model predictions for $\eta =100$, with
$\xi= 0.5$ (dot-dash curve) and $\xi= 1$ (dashed curve).
(b) (lower panel) Model predictions for  $\eta=500$, with
$\xi= 0.5$ (dot-dash curve) and $\xi= 1$ (dashed curve).}
\end{figure}

\begin{figure}
\psfig{figure=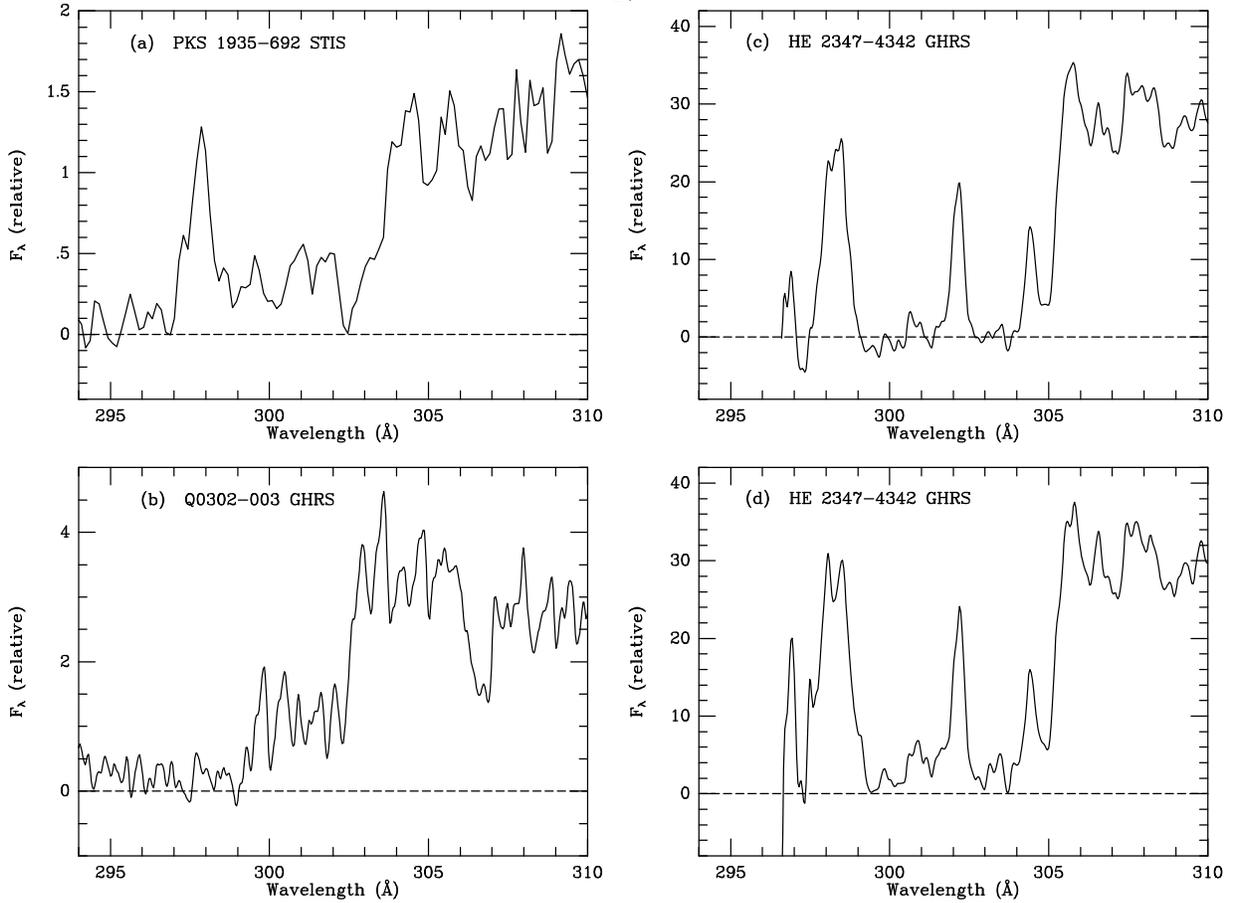,height=12truecm,width=16truecm,angle=270}
\caption[]{Recent spectra of the three ``HeII Gunn-Peterson quasars"
accessible to {\it HST} spectrographs, highlighting the regions near 
HeII, and with wavelengths in the  quasars' cosmic rest frames to
facilitate comparison. (a) (upper left panel) The new STIS 
spectrum of PKS~1935-692 presented in this work. 
(b) (lower left panel) The GHRS spectrum of Q0302-003 from HAR97, smoothed
to the GHRS spectral resolution. (c) (upper right panel) The GHRS spectrum
of HE~2347-4342 using the entire data set collected by Reimers et al. 
(1997); this spectrum is very similar to that discussed by Reimers et al.
but has been smoothed to the GHRS spectral resolution to facilitate
comparison with the other quasars. (d) (lower right panel) An alternate 
re-reduction of the GHRS spectrum of HE~2347-4342 that includes only a 
``low-noise"  (but potentially biased) subset of the data.
For all three quasars a correlation of HeII absorption features 
with HI Lyman-$\alpha$ clouds has now been demonstrated (e.g., the
HeII breaks themselves, strong voids in PKS~1935-692 and
HE~2347-4342, and other detailed absorption structure). 
A ``proximity shelf" and perhaps non-zero flux far from the
QSO redshift (blueward of the shelf) are seen in both
PKS~1935-692 and Q0302-003. (The alternate reduction for HE~2347-4342 
shown in the lower right panel potentially raises the possibility 
of a proximity shelf there as well.) Evidence for additional HeII
absorbing gas with $\Omega _g \sim10^{-2}$, that is not detected in HI, is seen 
in all three quasars.}
\end{figure}

\clearpage
\end{document}